# Comparison Of The Consumption Of Resources Between HTTP And SIP

Ravonimanantsoa N.Manda-Vy[1, a]**,** Randriamitantsoa P.Auguste [2,b]

[1]VS99bis Andranovory ambolokandrina Tana(101),Madagascar

[2]ESPA Université d'anatananarivo, Madagascar

[a]ndaohialy@blueline.mg, [b]rpa@freenet.mg,

**Abstract.** Currently, the development of research around VoIP experience a tremendous growth. In the community of open source Asterisk represents a reliable alternative for a lower cost solution. In this same community as the SIP protocol is a supplement to the more asterisk PBX. to share the benefits claimed by proponents of free software co-existence with other Asterisk server is not yet proven. In this context this paper we show a comparison of the use of simplified resource material for the apache server using the HTTP protocol and server that uses the asterisk SIP.

**Intorduction**

The protocol which dominates the current Internet infrastructure is http. But technological convergence over IP [1][2][3]networks leads us to see a large deployment of voice over IP on a global scale thanks to the many proprietary PABX. VoIP in general is from a server and most used in the free software world and the asterisk is most widely used protocol is SIP with Asterisk. But to get a better performance for both protocol and he needed to see what happens in the server ie their resource requirement and more specifically the level of memory consumption. To do so we'll at first briefly describe the environment of our experience, in a second you'll see the experience itself that is the course of the experiment and finally we will deduct from our experience a assumption on our experience.

**Methodology**

**Material .**
 The server:
As a server we used a PC dell optiplex GX 110 with a Pentium III processor and a minimum memory 256Mo
 The client:we used a PC dell optiplex GX 110 with a Pentium III processor and a minimum memory 128Mo
**Software:**
We used the operating system Debian GNU / Linux[4] that is a computer operating system composed of software packages released as free and open source software especially under the GNU General Public License and other free software licenses.The primary form, Debian GNU/Linux, which uses the Linux kernel and GNU OS tools,[4] is a popular and influential Linux distribution.It is distributed with access to repositories containing thousands of software packages ready for installation and use.
For the web server we used Apache. Apache is web server software notable for playing a key role in the initial growth of the World Wide Web.In 2009 it became the first web server software to surpass the 100 million website milestone.Apache was the first viable alternative to the Netscape Communications Corporation web server.
For the server we used Asterisk VoIP Asterisk is a software implementation of a telephone private branch exchange (PBX); it was created in 1999 by Mark Spencer of Digium. Like any PBX, it allows attached telephones to make calls to one another, and to connect to other telephone services including the public switched telephone network (PSTN) and Voice over Internet Protocol (VoIP) services. Its name comes from the asterisk symbol, "*".

Asterisk is released under a dual license model, using the GNU General Public License (GPL) as a free software license and a proprietary software license to permit licensees to distribute proprietary, unpublished system components.

Originally designed for Linux, Asterisk now also runs on a variety of different operating systems including NetBSD, OpenBSD, FreeBSD, Mac OS X, and Solaris. A port to Microsoft Windows is known as AsteriskWin32.

And for the evaluation of the consumption in memory we used TOP. The top command is a system monitor tool that produces a frequently-updated list of processes. By default, the processes are ordered by percentage of CPU usage, with only the "top" CPU consumers shown. top shows how much processing power and memory are being used, as well as other information about the running processes. Some versions of top allow extensive customization of the display, such as choice of columns or sorting method.

top is useful for system administrators, as it shows which users and processes are consuming the most system resources at any given time.

**Conducting the experiment:**
At first we run the top command and evaluated the memory used by the operating system. Several parts of the memory is already used for reasons within the system and for different services such as the launch of all deamon. The following figure(Fig. 1) shows the state of our system during the initial phase of our experience.

In a second time a customer accesses the apache server in our system and we do the sampling with the top command(Fig. 2)
and after sampling it disconnects the client.

In the third time we launch a call with a softphone, first you do not pick and we launch the top command (Fig.3 )and then pick up and watch as it happens(Fig. 4)

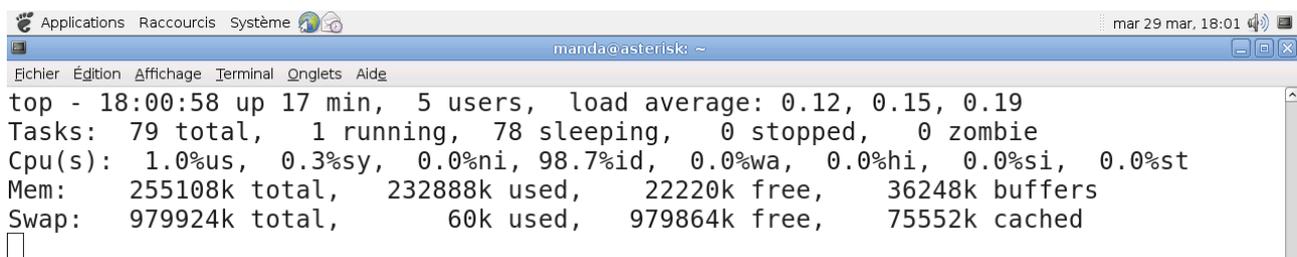

.Fig. 1

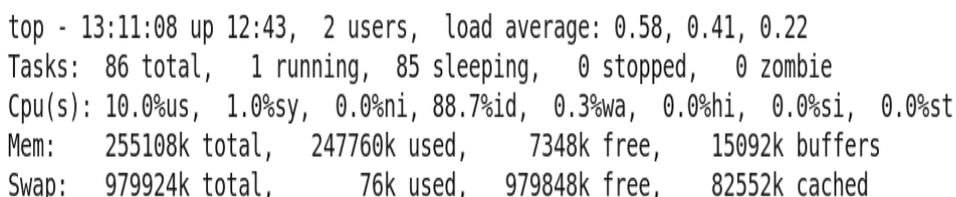

Fig. 2

```
 Applications  Raccourcis  Système                                                    mar 29 mar, 18:01
                              manda@asterisk: ~
Fichier Édition Affichage Terminal Onglets Aide
top - 18:01:19 up 18 min,  5 users,  load average: 0.09, 0.14, 0.18
Tasks:  79 total,   2 running,  77 sleeping,   0 stopped,   0 zombie
Cpu(s):  2.7%us,  1.0%sy,  0.0%ni, 96.3%id,  0.0%wa,  0.0%hi,  0.0%si,  0.0%st
Mem:    255108k total,    233128k used,    21980k free,    36292k buffers
Swap:   979924k total,        60k used,   979864k free,    75736k cached
```

Fig. 3

```
 Applications  Raccourcis  Système                                                    mar 29 mar, 18:01
                              manda@asterisk: ~
Fichier Édition Affichage Terminal Onglets Aide
top - 18:01:31 up 18 min,  5 users,  load average: 0.07, 0.13, 0.18
Tasks:  79 total,   2 running,  77 sleeping,   0 stopped,   0 zombie
Cpu(s):  4.0%us,  1.0%sy,  0.0%ni, 95.0%id,  0.0%wa,  0.0%hi,  0.0%si,  0.0%st
Mem:    255108k total,    233368k used,    21740k free,    36324k buffers
Swap:   979924k total,        60k used,   979864k free,    75928k cached
```

Fig. 4

**Results and commentary**
**Results:**

According to simulations, the phase initila mamoire free 22220ko is then connected when the client http free memory becomes 7348ko ie an http client consomne 14872ko memory and if we therefore multiplied the number of client server may stop responding.
 when initiating the call from a softphone free memory is 21980ko (this when the other end does not answer yet) then when it picks up the free memory becomes 21,740 which means that even if the customer lifts the receiver called the initiation of a SIP call consomne already resource at the server and the client called to consumption picks up in the memory increases.
ramarque:
the level of free memory 22220ko spent during the initial 21980ko during the launch of the appeal is to say that the memory occupied by a non-call pick up is 240ko. Then when the called party has picked up the level of memory is increased from 21980ko to 21740ko ie that Emmis's call with a second client occupies the same memory that the caller 240ko so if a call is has 480ko memory used and this we can deduce that the memory used(Mu) during a call 2nK Sip it with K = 240ko

$$Mu = 2nK \qquad (1)$$

**conclusion**
We have presented experimental results obtained during the simulation of a VoIP call and also the comparison with the http protocol. The different figures and results showed us that the http protocol consomne lot of memory compared to SIP, but a call sip Equival has a double connection to the server. It should be noted that the increase of the call can sip unanswered generated also a saturation of the mathematical model proposed mémoire.Le would then determine the number of calls made in relation to the size of memory used by the server.